\documentclass[preprint,aps,nofootinbib,preprintnumbers,amsmath,amssymb,11pt]{revtex4}
\usepackage{epsfig}
\graphicspath{ {images/} }
\usepackage{ytableau}
\usepackage{mwe}
\usepackage{simpler-wick}
\usepackage{amsmath}
\usepackage{multirow}
\usepackage{slashed}
\usepackage{xcolor}
\usepackage{amssymb}
\usepackage{color}
\definecolor{darkblue}{rgb}{0.,0.,0.4}
\definecolor{darkred}{rgb}{0.5,0.,0.}
\definecolor{BlueViolet}{RGB}{138,43,226}
\definecolor{SkyBlue}{RGB}{30,144,255}
\definecolor{DarkGreen}{RGB}{0,100,0}
\usepackage[colorlinks=true,linkcolor=blue,citecolor=blue]{hyperref}
\usepackage{amsfonts}
\usepackage{graphicx}
\usepackage{dcolumn}
\usepackage{bm}
\usepackage{natbib}

\newcommand{\be}{\begin{equation}}
\newcommand{\ee}{\end{equation}}
\newcommand{\bea}{\begin{eqnarray}}
\newcommand{\eea}{\end{eqnarray}}

\begin{document}


\title{From O(3) to Cubic CFT: Conformal Perturbation and the Large Charge Sector}

\author{\vspace*{0.5 cm}Junchen Rong}
\email{junchenrong@ihes.fr}
\affiliation{\vspace*{0.5 cm}Institut des Hautes \'Etudes Scientifiques, 91440 Bures-sur-Yvette, France}

\author{Ning Su}
\email{suning1985@gmail.com}
\affiliation{\vspace*{0.5 cm}Department of Physics, University of Pisa, I-56127 Pisa, Italy\\
and\\
Walter Burke Institute for Theoretical Physics, Caltech, Pasadena, California 91125, USA
\vspace*{2 cm}
}

\begin{abstract}
\vspace*{1 cm}
The Cubic CFT can be understood as the O(3) invariant CFT perturbed by a slightly relevant operator. 
In this paper, we use conformal perturbation theory together with the conformal data of the O(3) vector model to compute the anomalous dimension of scalar bilinear operators of the Cubic CFT.
When the $Z_2$ symmetry that flips the signs of $\phi_i$ is gauged, the Cubic model describes a certain phase transition of a quantum dimer model. The scalar bilinear operators are the order parameters of this phase transition. Based on the conformal data of the O(3) CFT, we determine the correction to the critical exponent as $\eta_{*}^{Cubic}-\eta_{*}^{O(3)}\approx -0.0215(49)$.
The O(3) data is obtained using the numerical conformal bootstrap method to study all four-point correlators involving the four operators: $v=\phi_i$, $s=\sum_i \phi_i\phi_i$ and the leading scalar operators with O(3) isospin $j=2$ and 4.
According to large charge effective theory, the leading operator with charge $Q$ has scaling dimension $\Delta_{Q}=c_{3/2} Q^{3/2}+c_{1/2}Q^{1/2}$. We find a good match with this prediction up to isospin $j=6$ for spin 0 and 2 and measured the coefficients $c_{3/2}$ and $c_{1/2}$.

\end{abstract}

\vspace*{3 cm}

\maketitle

\def\thesection{\arabic{section}}
\def\thesubsection{\arabic{section}.\arabic{subsection}}
\numberwithin{equation}{section}
\newpage

\tableofcontents

\section{Introduction}

The $O(3)$ vector model deformed by the cubic anisotropic term,
\be\label{cubicaction}
S=\int dt dx^2 \sum_{i=1}^3(\partial_{\mu} \phi_i )^2+ r \sum_{i=1}^3 \phi_i^2 + u (\sum_{i=1}^3 \phi_i\phi_i)^2 +v_4 \sum_{i=1}^3 (\phi_i)^4, 
\ee
was first proposed in \cite{aharony1973critical} to describe the crystal structural phase transition on perovskite materials. 
Over the years, the model has attracted the attention of physicists from different backgrounds.
In particular, an interesting question is to identify the most stable fixed point of \eqref{cubicaction} under renormalization group flow. 
The question was under debate for many years (as was reviewed in  \cite{Aharony:2022ajv}) until recently. 
The lattice Monte Carlo simulation result in  \cite{PhysRevB.84.125136,Hasenbusch:2022zur,Hasenbusch:2023fmn}, the perturbative quantum field thory \cite{KLEINERT1995284,PhysRevB.61.15136,Adzhemyan:2019gvv} and the conformal bootstrap result in \cite{Chester:2020iyt} all claim that the Cubic fixed point with non-vanishing coupling $v_4$ (instead of the $O(3)$ invariant fixed point) is the most stable one. 
The stability of the fixed points is closely related to the scaling dimension of the operator that introduces the cubic anisotropy, which belongs to the $j=4$ irreducible representation of O(3). 
The conformal bootstrap result \cite{Chester:2020iyt} provides rigorous proof that $\Delta_O <3$ at the $O(3)$ fixed point. This means that the Cubic model is in a new universality class.

Recently, the quantum Monte Carlo simulation in \cite{Yan:2022zwt} shows that the Cubic models also describe a second-order quantum phase transition of a quantum dimer model on a triangular lattice, building on earlier studies of the model such as in \cite{PhysRevLett.61.2376,PhysRevB.92.075141,moessner2001resonating,PhysRevB.99.165135,yan2021topological,Yan_2022}. 
The renormalization group flow structure of \eqref{cubicaction} plays an important role in understanding the phase diagram of the quantum dimer model.

Besides the above-mentioned connections to interesting phase transitions, the importance of the model \eqref{cubicaction} can be seen from a different angle. 
Consider $N$ scalar fields coupled together, to study possible conformal field theories, one can consider perturbation theory in $4-\epsilon$ expansion. 
Requiring the CFTs to have a single relevant operator, which corresponds to imposing the physical condition so that they correspond to critical points (rather than tri-critical points), the classification work of \cite{PhysRevB.10.892,PhysRevB.31.7171,Rong:2023xhz} shows that the list of such CFTs is very short, in particular when $N$ is small. 
If one chooses to extrapolate the result to $2+1$ dimensions, this means scalar universality classes that can be reached without fine-tuning are very limited.
The Cubic CFTs are valuable members of such a list.

Ever since the seminal work of \cite{Rattazzi:2008pe}, the conformal bootstrap method \cite{Poland:2018epd} has become one of the most successful methods in studying conformal field theories in 2+1 dimensions \cite{El-Showk:2012cjh,El-Showk:2014dwa,Kos:2014bka,Kos:2013tga,Kos:2015mba,Chester:2020iyt,Chester:2019ifh,Rong:2018okz,Atanasov:2018kqw,Kos:2016ysd}. 
However, even though there were some attempts \cite{Rong:2017cow,Stergiou:2018gjj,Kousvos:2018rhl}, the direct isolation of Cubic CFTs and the precise determination of their critical exponents remain elusive in numerical conformal bootstrap. 
Partially motivated by this goal, the model has been revisited recently using various perturbative methods \cite{Antipin:2019vdg,Bednyakov:2023lfj,Binder:2021vep}.

In this paper, we will use conformal perturbation theory together with the conformal data of the O(3) vector model to calculate the scaling dimensions of scalar bi-linear operators in the Cubic CFT. 
These operators are in fact the order parameter of the phase transition of the quantum dimer model \cite{Yan:2022zwt,Ran:2023wju}.
More precisely speaking, the low energy effective action that describes the phase transition is given by \eqref{cubicaction} coupled to a $Z_2$ gauge theory.
The $Z_2$ symmetry that was gauged flips the sign of all three $\phi_i$'s. 
Since the $\phi_i$ operator is not gauge invariant, they can not be the order parameter of the lattice model. 
The order parameters of the phase transition are instead composite operators $\phi_1 \phi_2$, $\phi_2 \phi_3$, and $\phi_1 \phi_3$. 
Such $Z_2$ gauged CFTs are usually denoted with a ``*'' and have been studied numerically before, the O(2)* universality was observed in \cite{sachdev1999translational,huh2011vison,YCWang2018}, while the O(4)* universality was studied in \cite{moessner2001ising,PhysRevLett.86.1881,yan2021topological}. 
In anticipation of future computer or even experimental measurement, it is therefore interesting to calculate the critical exponents $$\eta_{*}=2 \Delta_{\phi_1\phi_2}-1.$$ 
In conformal perturbation theory, the difference of the critical exponents $\eta_{*}$ between the Cubic CFT and the O(3) CFT is the leading order in the perturbation parameter $\delta$, which is related to the scaling dimension of the leading scalar operator in the $j=4$ irreducible representation of O(3) by $\delta=3-\Delta_{t_4} \approx 0.01$ \cite{Chester:2020iyt,Hasenbusch:2022zur,Hasenbusch:2023fmn}. 
The $\eta_{*}$ difference also depends on the OPE coefficient ratio $\alpha_{ttt_4}/\alpha_{t_4t_4t_4}$. 
The symbol $T^{(j)}$ here denotes the leading scalar operator in the isospin $j$ irreducible representation of O(3).  
To calculate $\alpha_{t_4t_4t_4}$, we set up, for the first time, a bootstrap program involving all four point correlators of $v=\phi_i$, $s=\sum_i \phi_i\phi_i$, $t$ and $t_4$. 
Our result shows that the $\eta_{*}$ difference between the Cubic CFT and the O(3) CFT start at two decimal places,
\be\label{etadifference}
\eta_{*}^{Cubic}-\eta_{*}^{O(3)}\approx -0.0215(49).
\ee
Together with  current most precise determination \cite{Chester:2020iyt} of $\eta_{*}^{O(3)}=2\Delta_T-(d-2)=1.41908(64)$, we get $\eta_{*}^{Cubic}\approx 1.398$.
Even though it seems challenging to observe such a difference in experiments, the recent studies in \cite{Hasenbusch:2022zur,Hasenbusch:2023fmn} suggest that the two CFTs are indeed distinguishable in Monte Carlo simulations. (In particular, critical exponents $\nu$ of the two CFTs were shown to be different \cite{Hasenbusch:2023fmn}.)

Our numerical bootstrap setup allows us to access many operators with high $O(3)$ representations. This allows us to check against the prediction of the scaling dimension of the large charge operators predicted by the large quantum charge effective theory~\cite{Hellerman:2015nra,Monin:2016jmo,Alvarez-Gaume:2016vff,PhysRevLett.120.061603,PhysRevLett.123.051603,Alvarez-Gaume:2019biu}, which states 
\be\label{largeQ}
\Delta_{Q}=c_{3/2} Q^{3/2}+c_{1/2} Q^{1/2} -0.094 +\sqrt{\frac{l(l+1)}{2}}
\ee
Reading the spectra of operators from extremal functional, we get
\footnote{We fitted equation (\ref{largeQ}) with our spectral data. See later sections for details. The numbers in the brackets are standard deviations from the fitting. However, this analysis does not take into account the error of the spectrum itself. Therefore, the actual error bars are probably slightly larger.}

\begin{equation}\label{universalcoefficient}
    c_{1/2}=0.304(35),\quad c_{3/2}=0.315(7)
\end{equation}
which are universal for all models that realize the O(3) universality class.

\section{The Cubic CFT from conformal perturbation}

As discussed in \cite{Komargodski:2016auf,Behan:2017emf,Behan:2017dwr}, when a CFT is perturbed by  an operator,
\be
\int dx^2 g O(x),
\ee
one may consider the renormalization group flow induced by such as perturbation.
The operator $O$ satisfies the normalization 
\be\label{normalization}
\langle O(x_1) O(x_2)\rangle=\frac{1}{x_{12}^{2\Delta_O}}, 
\ee 
If $\Delta_O=3-\delta$, the beta function is given by 
\be\label{betafunction}
\beta(g)=\frac{dg}{d\mu}=-\delta g -\frac{1}{2} C_{OOO} S_{d-1} g^2+\mathcal{O}(g^3).
\ee
$C_{OOO}$ is the OPE coefficient, and $S_{d-1}$ is the volume of unit (d-1)-sphere. The RG flow has an Infra-red fixed point when $C_{OOO}>0$ and $\delta>0$. One can read out the scaling dimension of the $O$ operator in the new fixed point, which corresponds to the $\omega$ critical exponent
\be
\omega=\frac{\partial\beta(g)}{\partial g}\bigg|_{g=g_*}=\delta.
\ee
Notice such a relation was shown to hold very precisely for the Cubic/O(3) CFT pair in \cite{Hasenbusch:2023fmn}.
Take another operator $\phi(x)$ (different from $O(x)$), again satisfying the normalization \eqref{normalization}, its anomalous dimension is given by 
\be\label{anldim1}
\gamma_{\phi}=\frac{2  C_{\phi\phi O }}{C_{OOO}} \delta+\mathcal{O}(\delta^2).
\ee
Higher order terms of \eqref{betafunction} and \eqref{anldim1} depend on the renormalization scheme. 
If the leading order vanishes (due to symmetry reasons), the next order terms are also scheme-independent. 

The operators of a CFT with O(3) symmetry are normalized as 
\be
\langle O^{(\Delta,l,j_1)}{}_{\alpha}(x_1,\xi_1) O^{(\Delta,l,j_2)}_{\beta}(x_2,\xi_2)\rangle=\delta_{j_1,j_2}G^{(j)}_{\alpha\beta}\frac{(I_{\mu\nu}(x_{12})\xi_1^{\mu}\xi_2^{\nu})^l-trace}{x_{12}^{2\Delta}},\quad I_{\mu\nu}(x)=\eta_{\mu\nu}-2\frac{x_{\mu}x_{\nu}}{x^2}.
\ee
Here $\xi^{\mu}$ is auxiliary null vectors. $j$ labels the O(3) representation, while $l$ denote the spin of the operator.The matrix $G^{(j)}_{ab}$ is defined in \eqref{metric} through Clebsch–Gordan coefficients.

The operator product expansion of two scalar operators is defined as 
\be\label{OPEa}
O^{(j_1)}_{1,a}(x_1)\times O^{(j_2)}_{2,b}(x_2)\sim \sum_{j} \alpha_{123}^{(j_1,j_2,j_3)} X^{(j_1,j_2,j_3)}_{abe} G^{(j)ef} C_{\Delta,l,\mu_1\ldots\mu_l}(x_{12}^{\mu},\partial_{2,\mu}) O^{\Delta,l,\mu_1\ldots\mu_l}_{3,f}{}^{}(x_2).
\ee
This implies the 3pt function be 
\be
\langle O^{(j_1)}_{1,a}(x_1) O^{(j_2)}_{2,b}(x_2) O^{(\Delta,l,j)}_{3,c}(x_3,\xi_3)\rangle= \alpha_{123}^{(j_1,j_2,j_3)} 
X^{(j_1,j_2,j_3)}_{abc}
\frac{(Z^{\mu}\xi_{3,\mu})^{l}}{x_{12}^{\Delta_{1}+\Delta_{2}-\Delta_{3}}x_{23}^{\Delta_{2}+\Delta_{3}-\Delta_{1}}x_{13}^{\Delta_{1}+\Delta_{3}-\Delta_{2}}},
\ee
with 
\be
Z^{\mu}=\frac{x_{13}^{\mu}}{x_{13}^2}-\frac{x_{12}^{\mu}}{x_{12}^2}.
\ee
The invariant tensor
$X^{(r,s,t)}_{abc}$ is defined in \eqref{invtensor}, which is proportional to the Clebsch–Gordan coefficient. We have provided a file containing all the conformal data of the O(3) CFT available for our bootstrap calculation. The above definitions for two and three-point functions are equivalent to saying that our conformal block follows the convention of the first line of Table I in \cite{Poland:2018epd}. Among the operators of the O(3) CFT, there are two operators that are relevant for our conformal perturbation calculation. They are the lowest dimension scalar operators in the $j=2$ and $j=4$ channel, we will denote them as $t$ and $t_4$.

We deform the O(3) CFT by the operator 
\be\label{cubicphi4}
M(x)=\sqrt{\frac{5}{6}} Q^{ijkl}e^{(4)a}_{ijkl} (t_4)_{a}(x).
\ee
The tensor $e^{(4)a}_{ijkl}$ is defined in \eqref{basischange2}, with $a=-4,...4$ and $i,j,k,l=1,..3$. This is used to change the basis from rank-4 symmetric traceless tensor to $J^2, J_3$ eigenstates basis in which the CG coefficients are defined (and used in the bootstrap setup). The tensor 
\be
Q^{ijkl} =\bigg\{
\begin{array}{l}
 1, \quad \text{if $i=j=k=l$} \\
 0,\quad \text{otherwise.} \\
\end{array}
\ee
is invariant under the Cubic group, which indicates the direction along which the O(3) group is broken into its Cubic subgroup. The constant $\sqrt{5/6}$ in (\ref{cubicphi4}) is chosen such that $M$ satisfies the normalization condition (\ref{normalization}).

The $j=2$ irrep of O(3), when decomposed into irreps of the Cubic group is given by the following branching rule:
\be
(j=2)\longrightarrow w+t'.
\ee
The order parameter of the Cubic$^*$ is defined as
\be
W(x)=\frac{1}{\sqrt{2}} w^{ij}e^{(2)a}_{ij}t_a(x).
\ee
with $e^{(2)a}_{ij}$ defined in \eqref{basischange2}. The tensor 
\be
w^{ij}=\left(
\begin{array}{ccc}
 0 & 1 & 0 \\
 1 & 0 & 0 \\
 0 & 0 & 0 \\
\end{array}
\right)
\ee
belong to the ``w'' irrep. It is also the irreps that the quadratic fields $\{\phi^1\phi^2,\phi^1\phi^3,\phi^2\phi^3\}$ of \eqref{cubicaction} transforms in.  
We can also calculate the anomalous dimension of 
\be
T'(x)=\frac{1}{\sqrt{6}} t'^{ij}e^{(2)a}_{ij}t_a(x).
\ee
The tensor
\be
t'^{ij}=\left(
\begin{array}{ccc}
 1 & 0 & 0 \\
 0 & 1 & 0 \\
 0 & 0 & -2 \\
\end{array}
\right)
\ee
transforms $t'$ irrep of the Cubic group. In terms of quadratic fields, the two-dimensional irrep is spanned by  $(\phi^1)^2+(\phi^2)^2-2(\phi^3)^2$ and  $(\phi^2)^2+(\phi^3)^2-2(\phi^1)^2$. The scaling dimension of the $t'$ operator controls the crossover critical behavior from the Cubic universality class to the Ising universality class \cite{PhysRevLett.33.427,Pelissetto:2000ek}, which can be realized in the structural phase transition of perovskites \cite{muller1975behavior}.

Using the above definition, we get that 
\be
C_{WWM}=-\frac{1}{3}\sqrt{\frac{2}{15}}\alpha_{ttt_4},\quad \quad
C_{T'T'M}=\sqrt{\frac{1}{30}} \alpha_{ttt_4},
\ee
and 
\be
C_{MMM}=\frac{7}{3} \sqrt{\frac{2}{429}} \alpha_{t_4t_4t_4}
\ee
Clearly, the $\alpha_{t_4t_4t_4}$ OPE coefficients appear in the conformal block expansion of the four-point function of 
$
\langle t_4(x_1)t_4(x_2)t_4(x_3)t_4(x_4)\rangle.
$
To get this OPE using conformal bootstrap, we set up a bootstrap program involving all four point correlators of $\phi_i$, $S=\sum_i \phi_i\phi_i$, $t$ and $t_4$. This is a heavy calculation. We leave the numerical details of the calculation in Appendix \ref{numericsdetails}. 
The bootstrap calculation gives us the OPE ratio
\be\label{operatio}
\alpha_{ttt_4}/\alpha_{t_4t_4t_4}\approx 0.62(14).
\ee
From these results, we get that\footnote{We used $\delta$ from Table \ref{fixedvalues}. Note that our value for $\Delta_{t_4}$ is outside the error bar from Martin's Monte Carlo study \cite{Hasenbusch:2022zur}. Our value for $\Delta_{t_4}$ is based on the Extreme Functional Method (EFM) and it is possible it is not accurate. If we instead used Martin's value from $\delta=0.0142$ \cite{Hasenbusch:2022zur}, the corrections are
\be
\gamma_{W}=2\frac{C_{WWM}}{C_{MMM}}\delta^{\text{MC}}\approx -0.0135(30),\quad \text{and}\quad \gamma_{T'}=2\frac{C_{T'T'M}}{C_{MMM}}\delta^{\text{MC}}\approx 0.0202(46). 
\ee
}
\be
\gamma_{W}=2\frac{C_{WWM}}{C_{MMM}}\delta\approx -0.0107(24),\quad \text{and}\quad \gamma_{T'}=2\frac{C_{T'T'M}}{C_{MMM}}\delta\approx 0.0161(36). 
\ee
By plugging $\Delta_t=1.20954$ \cite{Chester:2020iyt}, we get\footnote{The six-loop computation results from \cite{Bednyakov:2023lfj} are $\Delta_W=1.207, \Delta_{T'}=1.204$ (Our $W$, $T'$ are $Z$, $X$ of \cite{Bednyakov:2023lfj} respectively). }
\be
\Delta_{W}\approx 1.1988(24),\quad \text{and}\quad \Delta_{T'}\approx 1.2256(36).
\ee
In terms of the critical exponents $\eta_{*}$, we get \eqref{etadifference}.

\section{Large charge operators of the O(3) CFT}\label{largecharge}
Our numerical bootstrap setup allows us to access many operators with high $O(3)$ representations. 
To be precise, for Lorentzian scalars with $l=even$, operators with $O(3)$ isospin up to $j=8$ appear in our bootstrap setups.\footnote{Scalar operator with $j=7$ can not appear. The operators in $t_4\times t_4$ OPE are odd spin.}
The scaling dimension of those operators can be estimated by choosing a feasible point in our setup and using the Extreme Functional Method (EFM) \cite{ElShowk:2012hu}
On the other hand, large quantum charge effective theory~\cite{Hellerman:2015nra,Monin:2016jmo,Alvarez-Gaume:2016vff,PhysRevLett.120.061603,PhysRevLett.123.051603,Alvarez-Gaume:2019biu} predicts the asymptotic behavior of such operators to be (\ref{largeQ}).

\begin{figure}[!h]
\centering
\includegraphics[width = 0.6\textwidth]{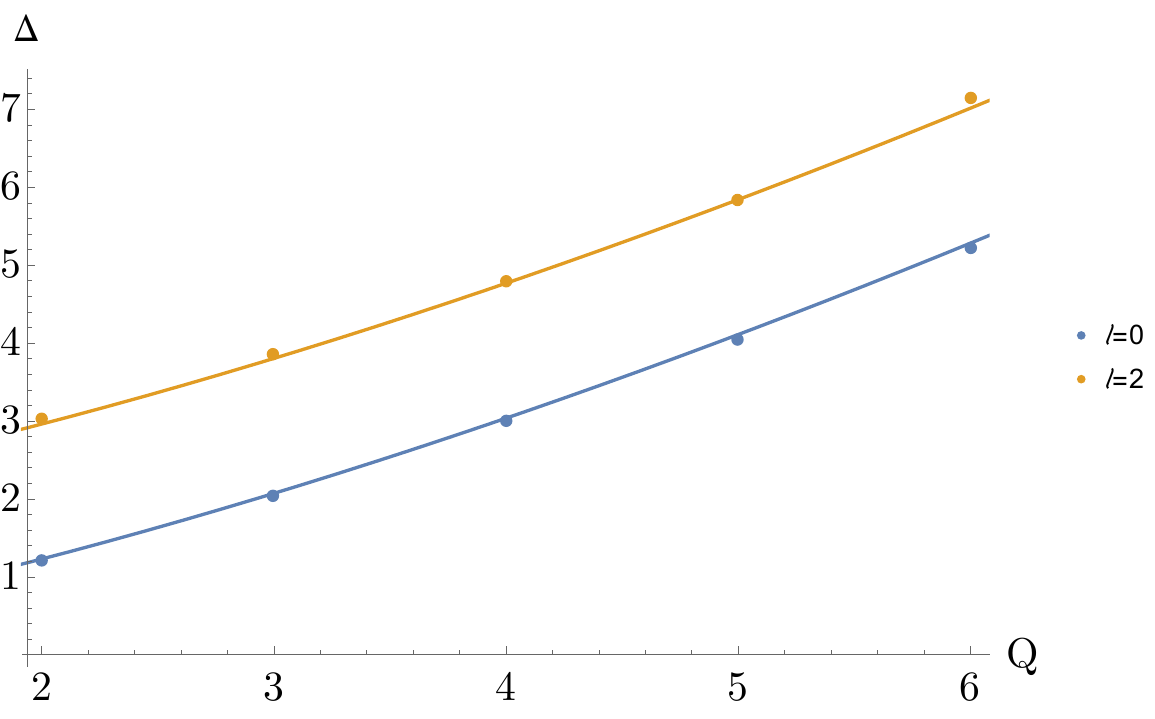}
\caption{\label{fig:large_charge_spectrum} The scaling dimension of large charge spin 0 and spin 2 operators, obtained using the Extremal Functional Method. The curves are obtained by fitting the spin-0 and spin 2 operators against the large charge formula $\Delta_{Q,l=0}=c_{3/2} Q^{3/2}+c_{1/2} Q^{1/2} -0.094$. 
}
\end{figure}

In the case of the O(3) model, one can take $Q=j$. Initially, it was expected that the above formula is valid for when $ 0\leq l\ll\sqrt{Q}$. 
The recent Monte Carlo simulation, however, shows that for O(2) and O(4) CFTs, the above asymptotic behavior works even when $Q\sim\mathcal{O}(1)$.
In Figure \ref{fig:large_charge_spectrum}, we plot our data against the large charge effective theory, for both $l=0$ and $l=2$ operators.\footnote{In our numerics, we observed certain fake operators below the expected value range. This is likely due to the sharing effect \cite{Simmons-Duffin:2016wlq,Liu:2020tpf}. They typically have smaller OPE coefficients in the EFM data than the actual operators. We didn't plot those operators in the figure.} Our results suggest that the large charge formula works even for small charge operators. 
We remark that this is the first time that \eqref{largeQ} has been tested for $l\neq 0$. 
For the charge 8 operators, our value is strongly affected by the sharing effect and unreliable. But it can be improved in the future by increasing $\Lambda$.

Fitting both spin 0 and spin 2 operators' dimensions by \eqref{largeQ}, we get \eqref{universalcoefficient}. From the figure we observed both sectors match well with the large charge expansion formula. 

\section{Discussion}
Using the conformal data obtained from our conformal bootstrap setup, we can also calculate the perturbation correction to the anomalous dimension of the critical exponents $\eta$ and $\nu$, corresponding to the operators $\phi_i$ and $\phi^2=\sum_i(\phi_i)^2$ of the Cubic CFT. 
Due to O(3) symmetry, the OPE coefficient $C_{\phi\phi t_4}$ and $C_{\phi^2\phi^2 t_4}$ both vanish. 
The conformal perturbation corrections to the scaling dimension of $\phi$ and $\phi^2$  start at order $\mathcal{O}(\delta^2)$, which should also be very small. 
This was first observed in an earlier field theoretical calculation \cite{calabrese2003randomly}, and confirmed using Monte Carlo simulation recently \cite{Hasenbusch:2022zur}. 
We will leave the conformal perturbation calculation of these critical exponents for future projects. 

In Section \ref{largecharge}, we see that large charge perturbation theory and the numerical bootstrap compensate each other, the former is valid for large charge operators while the latter is more accurate for small charges. 
It would be interesting to construct a hybrid bootstrap scheme to utilize the analytic information from large charge perturbation, similar to \cite{Su:2022xnj}. It would be great if one could directly bootstrap the Cubic CFT and obtain precise Cubic CFT data non-perturbatively. A promising approach was suggested in \cite{StefanosSaoPaoloTalk}.

\begin{acknowledgments}
We thank Joao Penedones for participating in the early stages of this work. We thank Johan Henriksson, Yinchen He, Junyu Liu, Luca Delacretaz, Alessandro Vichi, Gabriel Cuomo, and Slava Rychkov for insightful discussions. The manuscript was partially written during the Simons Bootstrap 2022 and 2023 annual meeting, for which we thank the University of Porto and ICTP South American Institute for Fundamental Research for its hospitality. The work of J.R. is supported by the Huawei Young Talents Programme at IHES. The work of N.S. was done mostly at the University of Pisa. This project has received funding from the European Research Council (ERC) under the European Union’s Horizon 2020 research and innovation program (grant agreement no. 758903). The computations in this paper were mainly run on the Symmetry cluster of Perimeter Institute. Research at Perimeter Institute is supported in part by the Government of Canada through the Department of Innovation, Science and Industry Canada and by the Province of Ontario through the Ministry of Colleges and Universities.

\end{acknowledgments}
\appendix
\section{Change of basis for O(3) group}
The \texttt{Autoboot} program works with Clebsch–Gordan (CG) coefficients, while in conformal perturbation theory, it is more convenient to work directly with the SO(3) indices $\{i,j,k\}$ carried by the SO(3) vectors irreps $x^i$. It is useful to know how to change the basis of these two conventions.
First, take the three vectors of the $j=1$ irrep of SO(3) to be $x_i$ with $i=1,2,3$. The $J_3$ eigenstates are 
\be\label{spin1states}
|+1 \rangle= \frac{x_1+i x_2 }{\sqrt{2}},\quad |0 \rangle=-x_3 ,\quad| -1 \rangle= -\frac{x_1 -i x_2}{\sqrt{2}}
\ee
States with higher $j$ can be calculated either by tensoring the above $j=1$ states using Clebsch–Gordan (CG) coefficients, which can be easily obtained by Mathematica command ``ClebschGordan[ ]'' or by acting creation operator $J_{-}=J_x-i J_y$ on the highest weight state 
\be
|j, m=j \rangle=(\otimes|+1 \rangle)^j=(\frac{x_1+i x_2}{\sqrt{2}})^j.
\ee
The above procedure writes all spin $j$ states in terms of the (symmetric) product of $j=1$ states. One can simply replace the $j=1$ states using \eqref{spin1states} to convert these states to polynomials.
From these polynomials, one can also construct symmetric trace-less tensors by 
\be\label{basischange1}
e^{(m)}_{ij}=\frac{1}{2!}\frac{\partial^2}{\partial x_i\partial x_j} |j=2, m \rangle.
\ee
This is a basis for rank-2 symmetric traceless tensors. Such basis tensors satisfy the following condition 
\be
e^{(m)}_{ij}e_{(n),ij}=\delta^{m}_{n},
\ee
where 
\be
e_{(n),ij}=(e^{(m)}_{ij})^*.
\ee
For the convenience of notation, we can also define a metric 
\be\label{basischange2}
e^{(a)}_{ij}e^{(b)}_{ij}=G^{(2)ab}=\left(
\begin{array}{ccccc}
 0 & 0 & 0 & 0 & 1 \\
 0 & 0 & 0 & -1 & 0 \\
 0 & 0 & 1 & 0 & 0 \\
 0 & -1 & 0 & 0 & 0 \\
 1 & 0 & 0 & 0 & 0 \\
\end{array}
\right).
\ee
Here $G^{(r)ab}$ is related to standard CG coefficients of SO(3) group by 
\be\label{metric}
G^{(r)}_{ab}=G^{(r)ab}=\sqrt{dim(r)}\left\{\frac{{a},{b}}{{r},{r}}\bigg|\frac{1}{id}\right\}_{so(3)}.
\ee
The standard CG coefficients can be obtained in  Mathematica through the command 
``ClebschGordan[ ]''.

A similar calculation can be performed to get a basis for rank-4 symmetric traceless tensors ($j$=4 states), which we denote as
\be
e^{(m)}_{ijkl}=\frac{1}{4!}\frac{\partial^4}{\partial x_i\partial x_j\partial x_k\partial x_l} |j=4,m \rangle.
\ee
The invariant tensors of O(3) appearing in the OPE \eqref{OPEa} are defined as 
\be\label{invtensor}
X^{(rst)}_{abc}=\left\langle \frac{a,b,c}{r,s,t}\right\rangle=\frac{1}{\sqrt{\dim  t}} \sum_{\bar{c}} 
\left\{\frac{a,b}{{r},{s}}\bigg|\frac{d}{{t}}\right\}_{so(3)}G^{(t)}_{cd}
\ee 

\section{Numerical bootstrap details}\label{numericsdetails}

The four-point correlator scalar operators satisfy the following crossing equation \cite{Rattazzi:2008pe},
\be
\wick{\langle \c1 \phi_1^{(r_1)} (x_1) \c1 \phi_2^{(r_2)} (x_2)\c1 \phi_3^{(r_3)} (x_3) \c1 \phi_4^{(r_4)} (x_4)}\rangle=\wick{\langle \c1 \phi_1^{(r_1)} (x_1) \c2 \phi_2^{(r_2)} (x_2)\c1 \phi_3^{(r_3)} (x_3) \c2 \phi_4^{(r_4)} (x_4)}\rangle,
\ee
where $r_i$ labels the $O(3)$ representation. We considered bootstrap equations from the following 17 correlators: $\langle ssvv \rangle$, $\langle stvv \rangle$, $\langle ttvv \rangle$, $\langle stt_4t_4 \rangle$, $\langle sst_4t_4 \rangle$, $\langle ttt_4t_4 \rangle$, $\langle st_4t_4t_4 \rangle$, $\langle tt_4t_4t_4 \rangle$, $\langle t_4t_4t_4t_4 \rangle$, $\langle sttt_4 \rangle$, $\langle tttt_4 \rangle$, $\langle tttt \rangle$, $\langle sstt \rangle$, $\langle sttt \rangle$, $\langle t_4t_4vv \rangle$, $\langle tt_4vv \rangle$, $\langle vvvv \rangle$. There are 82 independent equations from the crossing symmetry of these correlators. \footnote{As a comparison, using the same counting standard, the O(2) $v,s,t$ system has 22 equations \cite{Chester:2019ifh}, the O(3) $v,s,t$ system has 28 equations \cite{Chester:2020iyt}, the $O(5)$ $v,s,t$ system has 29 equations \cite{Chester:2023njo}, and the Potts $\sigma,\epsilon,\sigma'$ system has 39 equations \cite{Chester:2022hzt}.} The bootstrap equation can be collectively written as
\be
\sum_{r} \sum_{O : r} \overrightarrow{\lambda_r} \cdot V_{\Delta_{\mathcal{O}}, l_{\mathcal{O}}}(u, v) \cdot \overrightarrow{\lambda_r} = 0
\ee
where $O:r$ labels exchanged operators in the O(3) representation $r$, and $\overrightarrow{\lambda_r}$ are the vector of OPE coefficients of the form external-external-exchange. $V_{\Delta_{\mathcal{O}}, l_{\mathcal{O}}}(u, v)$ are a 82 dimension crossing vector whose entries are matrices that contract with $\overrightarrow{\lambda_r}$. We used \texttt{Autoboot} \cite{Go:2019lke,Go:2020ahx} to generate the bootstrap equations, and rewrote them as crossing vectors, which can be found in the attached file. 

Following the standard numerical conformal bootstrap approach \cite{Simmons-Duffin:2015qma}, we translate the bootstrap equation to semi-definite programs (SDP) that look for a linear functional $\alpha$ satisfying the positivity assumptions:
\bea
&&\alpha[V^{(r)}_{\Delta_{O},l_{O}}(u,v)]>0 \text{  for  } \Delta \ge \Delta^{(r)}_\text{gap}, \nonumber
\eea
where $r$ labels all $O(3)$ representations appearing in the OPE and $V^{(r)}$ are the crossing vectors. 

We put non-trivial gaps $\Delta^{(r)}_\text{gap}$ in various sectors, which are summarized in Table \ref{gaps}. Certain gaps are necessary for the mixed correlator system to be non-trivial --- without them, certain components of the functional $\alpha$ will be 0, a phenomenon observed in many mixed correlator bootstrap problems \cite{NingL4}.\footnote{One might wonder what the minimal set of gaps is that doesn't lead to a trivial mixed correlator bootstrap. However, we haven't tested this question carefully.} The values for the gaps are chosen based on the spectrum from \cite{Chester:2020iyt}, and the large charge expansion \cite{Banerjee:2017fcx,Banerjee:2019jpw}: we put relatively mild gaps with respect to the known values. The $\Lambda=43$ EFM data are from the mean values of the EFM spectrum obtained at various feasible points in the computation of $C_T, C_J$ in \cite{Chester:2020iyt}, and error bars are derived from the maximum and minimum of those values.\footnote{We thank Junyu Liu for helping to collect these data.} The error bars are not rigorous. For all other sectors, the gap is set to be the unitarity bound with a small twist gap $10^{-7}$, similar to the treatment in \cite{Chester:2019ifh}.

The SDP depends on the following parameters: $\Delta_v$, $\Delta_s$, $\Delta_{t}$, $\Delta_{t_4}$, $r_{tts}$, $r_{ttt}$, $r_{sss}$, $r_{vvt}$, $r_{t_4t_4s}$, $r_{t_4t_4t}$, $r_{t_4t_4t_4}$, $r_{ttt_4}$, where $r_{O_1O_2O_3}=\lambda_{O_1O_2O_3}/\lambda_{vvs}$ is the ratio of OPE coefficients.\footnote{
The convention of OPE coefficients in this paper (ab) is related to the convention of \cite{Chester:2020iyt} (CLLPSSV) by
\be
\left( \lambda_{sss}^{\text{CLLPSSV}}, \lambda_{tts}^{\text{CLLPSSV}}, \lambda_{vv t}^{\text{CLLPSSV}}, \lambda_{vv s}^{\text{CLLPSSV}}, \lambda_{ttt}^{\text{CLLPSSV}} \right) = \left( \lambda_{sss}^{\text{ab}}, \frac{1}{\sqrt{5}} \lambda_{tts}^{\text{ab}}, \frac{1}{\sqrt{10}} \lambda_{vv t}^{\text{ab}}, \frac{1}{\sqrt{3}} \lambda_{vv s}^{\text{ab}}, \sqrt{\frac{6}{35}} \lambda_{ttt}^{\text{ab}} \right).
\ee
} The resulting SDPs are large-scale. We compute the SDP at $\Lambda=19$. 

\begin{table}[h]
\begin{tabular}{|l|c|c|}
\hline
sector & $\Delta^{(r)}_\text{gap}$ & $\Lambda=43$ EFM data \cite{Chester:2020iyt} \\ \hline
$\text{v[1,-1] spin 0}$ & 4.9 & 5.003(15) \\ \hline
$\text{v[4,1] spin 0}$ & 6.4 & 6.573(60) \\ \hline
$\text{v[2,1] spin 0}$ & 3.4 & 3.559(03)\\ \hline
$\text{v[0,1] spin 0}$ & 3.6 & 3.767(36) \\ \hline
$\text{v[0,1] spin 2}$ & 4.6 & 4.738(37) \\ \hline
$\text{v[6,1] spin 0}$ & 5.0 & \\ \hline
$\text{v[8,1] spin 2}$ & 5.0 &  \\ \hline
\end{tabular}
\caption{The gap conditions we imposed in our setup. The $\text{v[j,z]}$ is the \texttt{Autoboot} notation for the $O(3)$ representation: $j$ denotes the O(3) isospin, and $z$ indicates the parity under the improper $Z_2$.
The last column provides the EFM spectrum data from the computation of $C_T, C_J$ in \cite{Chester:2020iyt}. The spin-0 $\text{v[6,1]}$ operator corresponds to the $Q$=6 operators in large charge expansion, for O(2) CFT and O(4) CFT, its scaling dimensions was $5.509(7)$ and $5.069(7)$ respectively, as measure using Monte Carlo simulation\cite{Banerjee:2019jpw,Banerjee:2017fcx}. Inspired by these results, $\Delta_{Q=4}>5.0$ for O(3) CFT is a safe gap assumption. The spin-2 $\text{v[8,1]}$ gap is also inspired by the large Q expansion.  
}
\label{gaps}
\end{table}

The parameters $\Delta_v, \Delta_s, \Delta_{t}, \Delta_{t_4}, r_{tts}, r_{ttt}, r_{sss}, r_{vvt}$ can be access by bootstrapping the O(3) correlators of $v, s, t$, which was done at $\Lambda=43$ in \cite{Chester:2020iyt}. The values from \cite{Chester:2020iyt} are likely much more accurate than our setup (the $v, s, t, t_4$ system) at $\Lambda=19$, since the main constraining power for those quantities comes from the $v, s, t$ system. Therefore we fixed those parameters to be the values summarized in Table \ref{fixedvalues}.

\begin{table}[h]
\begin{tabular}{|l|c|c|}
\hline
CFT data & Fixed value & Error bar from  \cite{Chester:2020iyt} \\ \hline
$\Delta_v$ &   0.5189415 & 0.518942(51) \\ \hline
$\Delta_s$ &   1.5949410 & 1.59489(59) \\ \hline
$\Delta_{t}$ & 1.2095570 & 1.20954(23) \\ \hline
$\Delta_{t_4}$ & 2.9886594 & (2.98640, 2.99056)\\ \hline
$r_{tts}$ & 2.4218778 & 2.42182(68) \\ \hline
$r_{ttt}$ & 3.9886334 & 3.98855(85) \\ \hline
$r_{sss}$ & 0.5569223 & 0.5567(11) \\ \hline
$r_{vvt}$ & 3.0455344 & 3.04548(42) \\ \hline
\end{tabular}
\caption{Parameters fixed in our bootstrap setup, that is, as inputs. In the second column, OPE ratios are the mean values of the $\Lambda=43$ feasible points in Table 6 of \cite{Chester:2020iyt}. The scaling dimensions are chosen by hand, but they fall within the error bars. In the third column, the data for $v, s, t$, and OPE coefficients are sourced from Eq. (20) and Eq. (24) of \cite{Chester:2020iyt} (converted to our convention). The upper bound of $\Delta_{t_4}$ is the rigorous bound from \cite{Chester:2020iyt}, while the lower value is derived from the EFM data using the same procedure as in Table \ref{gaps}.}\label{fixedvalues}
\end{table}

With the above parameters fixed, we scanned the parameters $r_{t_4t_4s}, r_{t_4t_4t}, r_{t_4t_4t_4}, r_{ttt_4}$ in our setup. This computation was done using the skydiving algorithm \cite{Liu:2023elz} and the framework software \texttt{simpleboot} \cite{simpleboot}. The parameters we used for the \texttt{skydive} program are the same as in the Table V of \cite{Chester:2023njo}. We performed three computations: (1) minimize the ratio $\lambda_{ttt_4}/\lambda_{t_4t_4t_4}$ to get the optimal point $p_\text{min}$; (2) maximize the ratio $\lambda_{ttt_4}/\lambda_{t_4t_4t_4}$ to get the optimal point $p_\text{max}$; (3) compute the EFM spectra (using \texttt{spectrum.py} \cite{Simmons-Duffin:2016wlq}) at $p_\text{min}$, $p_\text{max}$, and $p_0$, where $p_0$ is manually chosen roughly be at midpoint between $p_\text{min}$, $p_\text{max}$. 
\be
p_0 : (r_{t_4t_4s}, r_{t_4t_4t}, r_{t_4t_4t_4}, r_{ttt_4})=(6.03406111, 6.63716481, 9.00980854, 5.57430820)
\ee
The result of first two computations are summarized in Table \ref{result}, from which we conclude that the ratio $\lambda_{ttt_4}/\lambda_{t_4t_4t_4}=0.62(14)$.\footnote{Our error bar is not rigorous. For a rigorous error bar, we should scan over all parameters without fixing the values in Table \ref{fixedvalues}. But at $\Lambda=19$, this will be much less accurate, while at $\Lambda=43$, the SDP is too large and far beyond the computational capabilities of current-generation hardware. So we chose the compromised way to conduct our computation. An possible direction for future exploration is that we may assign a higher derivative order to correlators that only involve ${v,s}$ and a lower derivative order to other correlators.} 

The spectrum data at $p_0$ can be found in the attached file. We extracted the scaling dimensions of large charge operators from the EFM data at $p_\text{min}$, $p_\text{max}, p_0$.  We then averaged these dimensions and used the mean data for fitting and plotting in section \ref{largecharge}.

\begin{table}[h]
\begin{tabular}{|l|l|l|l|}
\hline
Computations & $r_{ttt_4}$ & $r_{t_4t_4t_4}$ &  $\lambda_{ttt_4}/\lambda_{t_4t_4t_4}$ \\ \hline
minimize ratio & 5.5760499378 & 11.6774329037 & 0.477506484838 \\ \hline
maximize ratio & 5.5762754523 & 7.34169704384 & 0.759534943906 \\ \hline
\end{tabular}
\caption{Skydive computation results}\label{result}
\end{table}

For those interested in the performance of these computations, we briefly summarize some key statistics here. For each computation, we used 4 nodes, and each node has 40 CPU cores. The first SDP in the skydiving computation takes about 2 days to finish. For the subsequent steps, each step (including the generation of SDP) takes about 30 minutes to 1 hour, depending on whether \texttt{skydive} decides to perform climbing steps. The computations (1), (2) take 128, and 222 steps, respectively. The entire computation takes about 10 days. The skydiving algorithm is essential for our computations because, without using the skydiving algorithm, we expect each step would take about 1 to 2 days to finish, and the entire computation could last for months or even years.

\bibliographystyle{JHEP}
\bibliography{main_v2.bib}

\end{document}